\documentclass{article}
\usepackage{longtable}

\usepackage[utf8]{inputenc}
\usepackage[T1]{fontenc}
\usepackage{graphicx}   % Required for inserting images
\usepackage{amsmath}    % For multi-line equations and math symbols
\usepackage{amssymb}    % For \mathbb and other symbols
\usepackage{bm}         % For bold math if needed
\usepackage{geometry}   % To manage margins
\usepackage{tabularx}

\usepackage{setspace}       % 行間設定用パッケージ
\begin{document}

% --- Title Section ---
\begin{center}
{\Large\bfseries Comparative visualization of knowledge structures using edge-difference graphs and network flow analysis: A case study of Wikipedia philosopher networks}
\vspace{1.5em}

{\large Towa Suda\textsuperscript{1*}, Susumu Hashimoto\textsuperscript{2}, Yuka Takedomi\textsuperscript{3}}
\vspace{1em}

{\small
\textsuperscript{1} Center for Digital Humanities and Social Sciences, Nagoya University, Nagoya, Aichi, Japan \\
\textsuperscript{2} Principles of Informatics Research Division, National Institute of Informatics, Chiyoda, Tokyo, Japan \\
\textsuperscript{3} HUMAI Center, ZEN University, Chuo, Tokyo, Japan
}
\vspace{1em}

{\small * Corresponding author: ts@nagoya-u.jp}
\end{center}
\vspace{1.5em}

% \newpage % arXivではアブストラクトを独立したページにせず、そのまま下に続けるのが一般的です（改ページを削除）

\begin{abstract}
Visualizing knowledge structures as graphs is common, but making them intuitively understandable remains challenging. Existing methods, such as macroscopic statistical metrics and whole-graph visualizations, often fail to capture local differences in conceptual relationships and suffer from severe visual clutter as networks grow large. To address these limitations, we propose a comparative visualization method that combines Edge-Difference Graphs with network flow analysis. The method first constructs Edge-Difference Graphs by extracting edges unique to each graph from graphs sharing a common node set, reducing redundancy while preserving the overall graph structure. It then identifies diverse paths between specific nodes by solving a minimum-cost maximum-flow problem. By incorporating a cost based on Adamic-Adar similarity, it penalizes routes that pass through generic hub concepts, enabling the extraction of contextually specific paths. We applied the method to networks of 20th-century French philosophers constructed from the French, German, English and Japanese editions of Wikipedia. The results reveal distinctive relational paths that reflect how each linguistic community receives and contextualizes these philosophers. This study provides a framework for the comparative analysis of large-scale knowledge structures and deepens our understanding of cultural and structural differences.
\end{abstract}

\section{Introduction}
Understanding a knowledge structure requires more than understanding individual concepts; it also requires recognizing the relationships among them. Prior research in knowledge representation and learning theory supports this view, suggesting that knowledge is not a collection of isolated concepts but a structured network. In particular, research on concept maps characterizes knowledge as a structured network of concepts connected by labeled relations and regards propositions composed of concept-relation-concept units as fundamental elements of meaning~\cite{novak2008}. Therefore, a deep understanding of a knowledge structure requires grasping its overall relational structure.

Visualizing knowledge structures as graphs is common for representing relational structure~\cite{hogan2021, fortunato2010}, but intuitively comparing their structural characteristics remains challenging. Comparing whole-graph statistical indicators, such as network density, is effective for quantifying macroscopic features, but fails to provide a concrete understanding of how local relationships differ~\cite{wills2020}. In addition, simply visualizing the entire graph has a major drawback: as the scale grows, the dense accumulation of nodes and edges causes visual clutter, which severely degrades readability and makes structural differences difficult to discern~\cite{herman2000, ghoniem2004}.
To address these challenges, this study proposes a novel comparative visualization method that combines edge-difference analysis and network flow optimization. By computing the set difference of edges across graphs sharing a common node set, we extract relationships unique to each structure, reducing redundancy while preserving the overall context. Then, by applying the minimum-cost maximum-flow algorithm with a cost function derived from Adamic-Adar similarity, we extract diverse paths between specific nodes while avoiding routes through high-degree hub nodes. As a case study, we applied this method to networks of 20th-century French philosophers constructed from multiple language editions of Wikipedia.

\section{Related works}

\subsection{Graph visualization of knowledge structures and the clutter problem}
Representing and analyzing knowledge as graph structures is foundational to modern information science, including data integration and reasoning~\cite{hogan2021}. As the theory of concept maps indicates, the process of acquiring expertise is synonymous with positioning new concepts within an existing knowledge network; thus, focusing on the topology and local connection patterns between nodes is essential for a deep understanding of knowledge structures~\cite{novak2008}. While community detection and other methods are widely used to extract structural features within a network~\cite{fortunato2010}, intuitively grasping the differences when comparing multiple knowledge structures remains challenging.

Macroscopic statistical metrics, such as network density and clustering coefficients, are useful for measuring overall graph similarity but cannot identify exactly which relationships between concepts differ~\cite{wills2020}. On the other hand, visualizing the entire graph suffers from severe visual clutter, where nodes and edges become densely packed as the data scale increases~\cite{herman2000}. Because this obscures fine structural differences, clutter reduction techniques like sampling and filtering~\cite{ellis2007}, as well as hierarchical graph reduction approaches like sparsification and coarsening~\cite{graphreduction2024}, have become indispensable.

\subsection{Edge-Difference Graphs and network flow analysis}
Visualizing graph differences enables the comparison of different graph structures while reducing visual clutter~\cite{differenceviews2025}. In our framework, by computing the set difference across multiple graphs and retaining only the unique edges, this approach isolates structural differences for focused analysis. It has been adopted to control cognitive load in visualizations that capture changes in dynamic network flow patterns~\cite{netostat2021} and track the dynamic evolution of academic disciplines~\cite{temporalgraph2006}.

While Edge-Difference Graphs isolate the edges unique to a given structure, interpreting these local differences within the overall relational context requires network flow analysis. The maximum flow problem enables the discovery of multiple independent contexts (edge-disjoint paths) connecting two nodes, and the minimum-cost maximum-flow formulation further enables the extraction of diverse paths under cost constraints~\cite{ahuja1993network}.

A persistent challenge in applying flow-based methods to knowledge graphs is the presence of ``hub nodes'', which are high-degree nodes that act as intermediaries between most node pairs, reducing structurally distinct paths to trivial routes through a few general concepts~\cite{conceptnet2019}. In the link prediction literature, the Adamic-Adar index~\cite{adamic2003} quantifies the specificity of shared neighbors by weighting them inversely by their degree, and has been used to penalize paths through high-degree nodes~\cite{phub2013}. To our knowledge, no existing study has combined edge-difference analysis with cost-penalized network flow to extract culturally specific paths from multilingual knowledge graphs.

\subsection{Multilingual networks and cultural differences in Wikipedia}
Wikipedia serves as an ideal corpus for comparing multilingual and multicultural knowledge structures~\cite{plosone2014}. Studies show that Wikipedia contains strong cultural bias, and the volume and focus of descriptions vary greatly by language even for the same topic~\cite{culturalbias2011}. In fact, each language edition contains a large amount of Cultural Context Content specific to its cultural sphere, creating significant information gaps between languages~\cite{frontiers2018}.

Particularly in philosopher networks, previous studies have shown that the importance of a subject and the contexts associated with it fluctuate significantly depending on the language edition (cultural sphere)~\cite{philosophers2025}. Our proposed method captures these cultural variations by removing general hubs from these multilingual networks, allowing us to extract the unique cultural contexts specific to each language.

\section{Materials and methods}

\subsection{Graph definitions}
Let $G_1^{\text{base}} = (V_1, E_1), G_2^{\text{base}} = (V_2, E_2), \dots, G_K^{\text{base}} = (V_K, E_K)$ be directed graphs representing the raw knowledge structures extracted from $K$ different language editions of Wikipedia. To enable a strict cross-lingual comparison, we first define a unified and common node set $V$ by taking the intersection of all individual node sets:
\begin{equation}
V = \bigcap_{k=1}^{K} V_k.
\end{equation}
Using this common node set, we construct the \emph{Filtered graph} $G_i^{\text{filt}} = (V, E_i^{\text{filt}})$ for each language $i$, where the edge set contains only the links between the mutually shared nodes:
\begin{equation}
E_i^{\text{filt}} = E_i \cap (V \times V).
\end{equation}

For each graph, we assign a cost to each directed edge $(u, v)$ based on the Adamic-Adar similarity~\cite{adamic2003}. For a directed graph $G = (V, E)$, let $N(u)$ denote the union of the in-neighbors and out-neighbors of node $u$ in $G$. The Adamic-Adar similarity between two nodes $u$ and $v$ is defined as:
\begin{equation}
AA(u, v) = \sum_{w \in N(u) \cap N(v)} \frac{1}{\log |N(w)|}.
\end{equation}
We then invert this similarity into an edge cost:
\begin{equation}
d_{AA}(u, v) = \frac{1}{1 + AA(u, v)}
\end{equation}
This cost is normalized to the range $(0, 1]$, reaching its maximum value of $1$ when $AA(u, v) = 0$. When two nodes share many low-degree neighbors, $AA(u, v)$ is large and the cost is low; when their shared neighbors are high-degree hub nodes, $AA(u, v)$ is small and the cost is high. In this way, paths through specialized nodes are favored over paths through general hubs.

We introduce the \emph{Edge-Difference Graph} to isolate edges unique to a single knowledge structure (Fig 1.). For a target graph $G_i^{\text{filt}}$, we define $G_i^{\text{diff}} = (V, E_i^{\text{diff}})$ by retaining only the edges absent from all other filtered graphs:
\begin{equation}
E_i^{\text{diff}} = E_i^{\text{filt}} \setminus \bigcup_{j \neq i} E_j^{\text{filt}}
\end{equation}

The resulting graph preserves the original node set $V$ but contains only the edges distinctive to $G_i$. In this study, we compute $AA(u, v)$ and $d_{AA}(u, v)$ on $G_i^{\text{diff}}$, so that the cost reflects only the unique relational structure of each language edition. Computing the Adamic–Adar index directly on the Edge-Difference Graph ensures that the edge costs are penalized based exclusively on the neighborhood structure unique to that specific language edition. This approach effectively penalizes ``local hub nodes''—entities that may not be prominent global hubs across all editions but act as generic connectors within the distinctive edge set of a single cultural sphere—thereby prioritizing paths formed by contextually specialized domestic relationships.

\begin{figure}[htbp]
  \centering
  \includegraphics[width=0.85\textwidth]{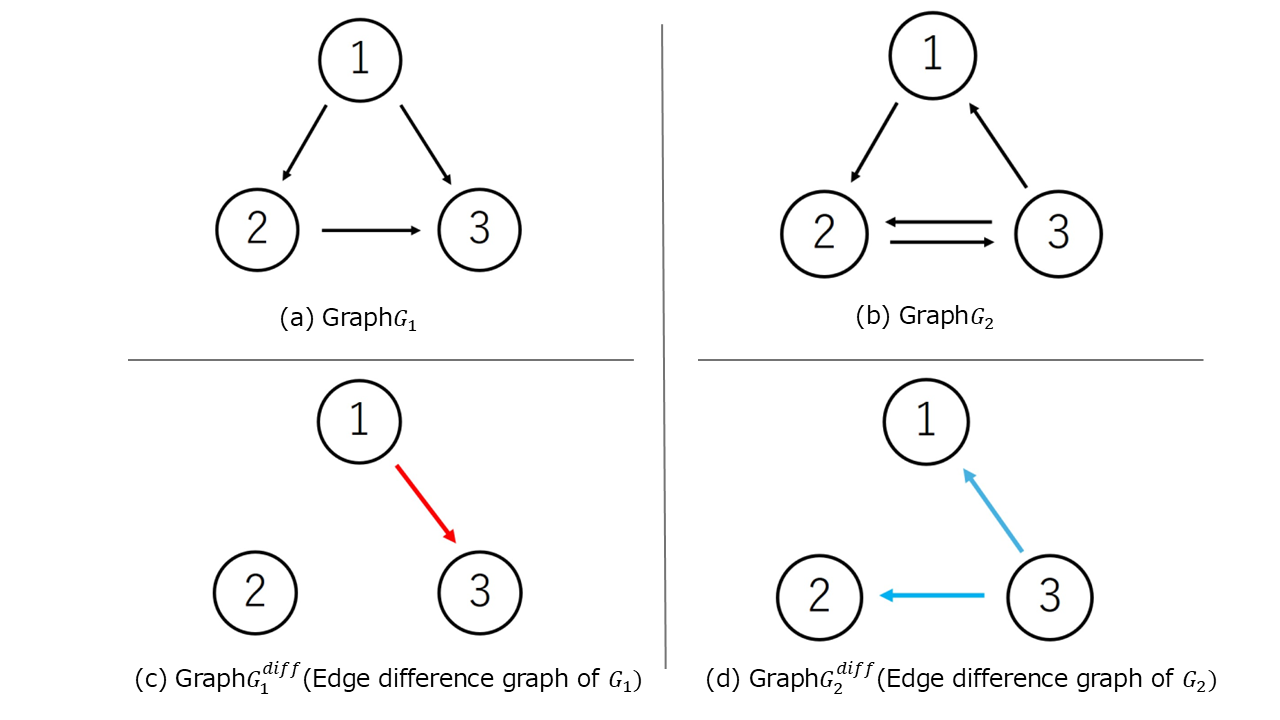}
  \caption{\textbf{Framework of Edge-Difference Graph generation.} (a) Target graph $G_1$ and (b) comparison graph $G_2$ sharing a common node set. (c) The Edge-Difference Graph $G_1^{\text{diff}}$ isolates the edge unique to $G_1$ (red arrow) by removing mutually shared links. (d) The Edge-Difference Graph $G_2^{\text{diff}}$ similarly isolates the edges unique to $G_2$ (blue arrows).}
  \label{fig:framework}
\end{figure}

\subsection{Network flow analysis for path extraction}
To discover diverse, low-cost paths between any two nodes in the Edge-Difference Graph, we apply the minimum-cost maximum-flow algorithm. Let $G_i^{\text{diff}} = (V, E_i^{\text{diff}})$ be an Edge-Difference Graph with a source node $s \in V$ and a target node $t \in V$. Each edge $(u, v) \in E_i^{\text{diff}}$ has a capacity $c(u, v)$ and a cost $d_{AA}(u, v)$. A flow $f: V \times V \to \mathbb{R}$ must satisfy the following constraints:
\begin{equation}
0 \le f(u, v) \le c(u, v) \quad \text{for all } (u, v) \in E_i^{\text{diff}}
\end{equation}
\begin{equation}
\sum_{u \in V} f(u, v) = \sum_{w \in V} f(v, w) \quad \text{for all } v \in V \setminus \{s, t\}
\end{equation}
The first condition ensures that the flow on each edge does not exceed its capacity and is non-negative, while the second guarantees flow conservation at all nodes except the source and target.

The total flow value from $s$ to $t$ is defined as
\[
B = \sum_{w \in V} f(s, w) - \sum_{w \in V} f(w, s).
\]
Let $B^*$ denote the value of a maximum flow from $s$ to $t$. The minimum-cost maximum-flow is the flow of value $B^*$ that minimizes the total cost
\[
\Pi = \sum_{(u, v) \in E_i^{\text{diff}}} f(u, v)\,d_{AA}(u, v).
\]

In our framework, we set the capacity of each edge uniformly to $c(u, v) = 1$. Under this setting, a maximum flow of value $B^*$ corresponds to $B^*$ edge-disjoint paths (where no edge is reused) from $s$ to $t$. Solving the minimum-cost maximum-flow problem then yields the set of $B^*$ edge-disjoint paths with the minimum total cost. We first compute $B^*$ using a standard maximum flow algorithm and then solve for the minimum-cost flow of value $B^*$ via the network simplex method. The resulting flow is decomposed into individual paths. Because $d_{AA}$ assigns high cost to edges whose shared neighbors are high-degree hub nodes, the extracted paths favor routes through specialized nodes (Fig 2.).

\begin{figure}[htbp]
  \centering
  \includegraphics[width=0.85\textwidth]{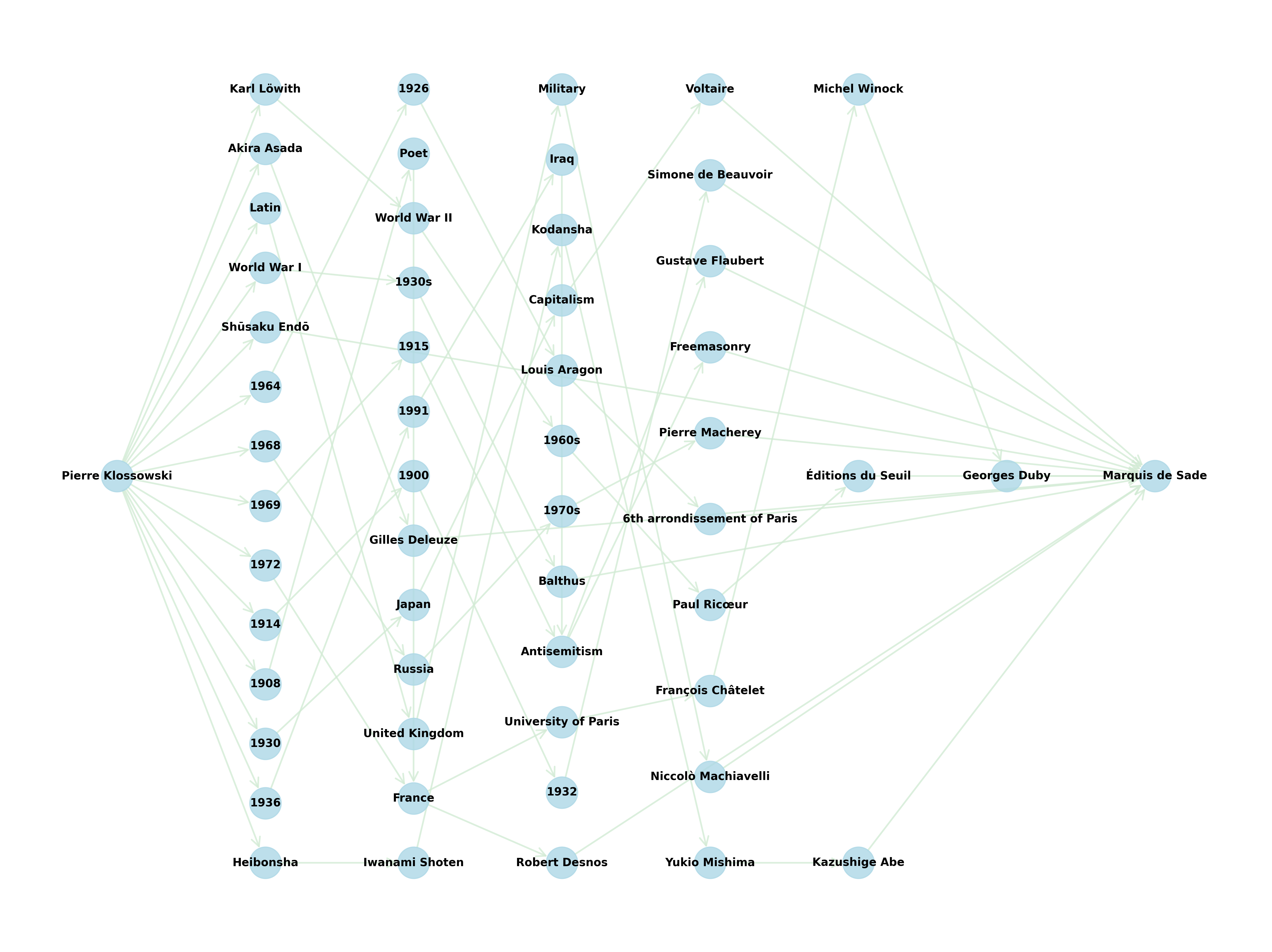}
  \caption{\textbf{Path extraction using minimum-cost maximum-flow on the Edge-Difference Graph.} Each edge has unit capacity and cost $d_{AA}(u, v)$. The algorithm extracts $B^*$ edge-disjoint paths from $s$ to $t$ that minimize the total cost, favoring routes through low-degree specialized nodes over high-degree hubs.}
  \label{fig:flow}
\end{figure}

For graph construction and network flow computation, we used Python (version 3.13) and NetworkX (version 3.5).

\subsection{Data collection and network construction}
The networks used in this study were constructed from internal links (wikilinks) in four language editions of Wikipedia (French, German, English and Japanese). Data were collected between October and November 2025. The construction process consisted of four steps.

\textbf{Step 1: Selection of initial nodes.}
The starting point for our data collection was a list of 131 philosophers featured in the 20th Century (I, II) chapters of the \textit{Dictionary of French Philosophy and Thought}~\cite{frenchdict1999}. Using Wikidata QIDs as keys, we identified the corresponding article titles in the French, English, German, and Japanese Wikipedia editions. Of the 131 philosophers, 130 had an article in at least one of the four language editions; the remaining individual was excluded. These 130 nodes constitute the ``initial node group''. Relying on QIDs ensures that entities with different article titles across languages---for example, ``Michel Foucault'' in French and its corresponding title in Japanese---are identified as the same entity.

\textbf{Step 2: Network expansion and edge definition.}
To incorporate individuals and concepts closely related to the initial philosophers, we expanded the network. First, we retrieved the wikitext for the 130 initial nodes across the four language editions and extracted all internal links within the main article namespace. Next, we queried the Wikidata API to obtain QIDs for these linked article titles. However, this process left some links without QIDs, an issue addressed in Step 3. The resulting node list comprised 19,249 unique nodes. Finally, we retrieved the wikitext for all 19,249 nodes across all target language editions and extracted their internal links. A directed edge was defined if and only if both the source and  target QID of a link were present in the unified node list.

\textbf{Step 3: Cross-lingual resolution of missing QIDs.}
Converting link strings into language-independent QIDs was a central challenge. API queries often fail to return QIDs due to redirects or orthographical variants. To maximize the retrieval rate, we implemented a cross-lingual resolution process. We referenced the top 10 Wikipedia language editions by edit count as of October 2025, which include our four target languages (English, German, French, Spanish, Russian, Italian, Japanese, Chinese, Polish, and Vietnamese). If a QID could not be retrieved for a link string in one language, we searched for that string across the other editions. If a matching page with a QID was found, we assigned that QID to the missing value.

\textbf{Step 4: Filtering for common nodes.}
Following the cross-lingual resolution, the initially constructed networks for each language edition had different numbers of nodes due to varying coverage across Wikipedia communities. To establish an identical baseline for subsequent edge-difference analysis, we extracted the intersection of the node sets across all four language editions. Nodes that were not mutually shared across all editions, along with their incident edges, were filtered out. This process yielded a final set of 8,213 common nodes, forming the Filtered graphs used as the basis for our comparative analysis.

\section{Results}

\subsection{Basic network statistics and structural characteristics}
The structural properties of the philosopher networks across the four language editions are summarized in Table~\ref{tab:stats}. In the raw dataset (Base), the French edition had the largest network, followed by the English, German, and Japanese editions in node count. However, regarding connection density, the French edition was the most densely connected, followed by the Japanese, English, and German editions in descending order.

A critical step in our analysis was the construction of the Filtered graph, which consists of 8,213 nodes common to all four language editions. Despite having the smallest node count in the Base graph (10,007), the Japanese edition exhibited the highest ``node commonality'', with approximately 82\% of its nodes being shared across all languages.

However, a significant divergence appears in the Edge-Difference Graphs, which contain only the edges unique to a specific language within the Filtered graph's node set. As shown in Table~\ref{tab:stats}, the Japanese edition has a substantial number of unique relationships (141,795 edges), resulting in an average degree of 34.53 for its Edge-Difference structure. This indicates that even when dealing with a common set of global entities, the Japanese Wikipedia constructs a distinct relational context.

\begin{table}[htbp]
\centering
\caption{Basic statistics of the philosopher networks (Base, Filtered, and Edge-Difference).}
\label{tab:stats}
\begin{tabular}{llrrr}
\hline
\textbf{Graph Type} & \textbf{Language} & \textbf{Nodes} & \textbf{Edges} & \textbf{Average Degree} \\ \hline
Base  & French & 16,373& 581,882& 71.08\\
 & German & 12,890& 269,588& 41.83\\
 & English & 14,907& 338,217& 45.38\\

 & Japanese & 10,007& 305,245& 61.01\\ \hline
Filtered  & French & 8,213& 330,587& 80.50\\
& German & 8,213& 180,672& 44.00\\
 & English & 8,213& 231,122& 56.28\\
 & Japanese & 8,213& 250,365& 60.97\\ \hline
Edge-Diff.  & French & 8,213& 197,043& 47.98\\
& German & 8,213& 89,880& 21.89\\
 & English & 8,213& 107,289& 26.13\\
 & Japanese & 8,213& 141,795& 34.53\\ \hline
\end{tabular}
\end{table}

\subsection{Path extraction and multilingual comparison}
To qualitatively interpret the structural differences identified in the statistical analysis, we applied the minimum-cost maximum-flow algorithm to the edge-difference graphs. Rather than arbitrarily selecting cases, we established an evaluation framework under two distinct contextual settings to test the methodological capability across different conceptual boundaries. Group 1 focuses on the immediate network surrounding a core initial node (Pierre Klossowski, Q203674) and its primary ideological reference points (Table~\ref{tab:group1}). Group 2 focuses on the network surrounding a major external boundary node (Gayatri Chakravorty Spivak, Q240851) and its theoretical intersections (Table~\ref{tab:group2}). 

Klossowski was selected for Group 1 to investigate how the framework captures the structural formation of an initial node's primary intellectual lineage. In intellectual history, figures like Friedrich Nietzsche and the Marquis de Sade—though outside the initial 130 core French nodes—served as foundational reference points for Klossowski's idiosyncratic philosophy. Extracting paths within this group enables us to evaluate how effectively the algorithm uncovers the unique local paths through which different linguistic editions reconstruct the historical and philosophical genealogy of a core figure. 

Conversely, Spivak was selected for Group 2 to test the framework's effectiveness on a network beyond the initial 130 core nodes. As a foundational post-colonial theorist who translated and adapted French post-structuralism, Spivak is connected to relevant figures such as Edward Said, Homi K. Bhabha, and Judith Butler, involving bidirectional links (both incoming and outgoing edges). Evaluating the paths within this surrounding network enables us to demonstrate the model's flexibility in capturing how different language editions reconstruct local critical contexts, even for entities outside the core set.

The analysis reveals that each language edition captures paths reflecting its unique cultural and historical context. In the German edition, paths often incorporate German-speaking thinkers such as Hannah Arendt and J\"urgen Habermas, or significant historical events like the Nazi seizure of power. Similarly, the English edition highlights Anglophone literary figures like Salman Rushdie and Rudyard Kipling. A quantitative overview further underscores the richness of the home-language data: for Group 1, the French edition yields the highest number of unique paths (35 paths for Klossowski to Nietzsche, Bataille, Gide, Lacan, and Foucault, and 31 paths to Sade), substantially outperforming the Japanese (21 paths) and English (9 paths)  editions. This density reflects a deeply saturated relational matrix within the French Wikipedia that persists even after removing shared global links (see Table 4 and Table 5 for qualitative path details).

Beyond the qualitative itineraries, the total number of edge-disjoint paths ($B^*$) itself serves as a crucial structural indicator of localized contextual connectivity. As recorded in the ``Rank/Total'' column of Table~\ref{tab:group1_paths} and Table~\ref{tab:multilingual_paths_full}, the volume of unique paths varies significantly across language editions. In Group 1, the trajectory from Klossowski to Nietzsche yields 40 independent paths in the German edition, followed by the French (35), Japanese (21), and English (9) editions. Similarly, in Group 2, the trajectory from Spivak to Jacques Derrida yields 19 independent paths in the German edition, followed by the English (16), French (14), and Japanese (4) editions. This variation demonstrates that the capacity for culture-specific narrative diversity is heavily dependent on the local topology surrounding specific entity pairs.

\begin{table}[htbp]
\centering
\caption{Target nodes for Group 1 (Klossowski-centered pairs).}
\label{tab:group1}
\begin{tabular}{ll}
\hline
\textbf{Target Philosopher} & \textbf{Wikidata QID} \\ \hline
Georges Bataille & Q207359\\
Jacques Lacan & Q169906\\
Michel Foucault & Q44272\\
Friedrich Nietzsche & Q9358\\
Andre Gide & Q47484 \\
Marquis de Sade & Q123867\\ \hline
\end{tabular}
\end{table}

\begin{table}[htbp]
\centering
\caption{Target nodes for Group 2 (Spivak-centered pairs).}
\label{tab:group2}
\begin{tabular}{ll}
\hline
\textbf{Target Philosopher} & \textbf{Wikidata QID} \\ \hline
Gilles Deleuze& Q184226\\
Edward Said & Q201538\\
Homi K. Bhabha & Q325741\\
Judith Butler & Q219368\\
Jacques Derrida & Q130631\\
Michel Foucault & Q44272\\ \hline
\end{tabular}
\end{table}

\begin{table}[!ht]
\centering
\onehalfspacing
\scriptsize
\caption{Comparison of characteristic paths for Group 1 (Klossowski-centered pairs).}
\label{tab:group1_paths}
\begin{tabularx}{\textwidth}{lll>{\raggedright\arraybackslash}X}
\hline
\textbf{Language} & \textbf{Source $\to$ Target} & \textbf{Rank/Total} & \textbf{Extracted Path} \\ \hline
French & Pierre Klossowski $\to$ Georges Bataille & 2 / 35 & Pierre Klossowski $\to$ 1935 $\to$ \textbf{\textit{Éditions Gallimard}} $\to$ \textbf{\textit{Les Temps modernes}} $\to$ Georges Bataille \\
French & Pierre Klossowski $\to$ Georges Bataille & 13 / 35 & Pierre Klossowski $\to$ \textbf{Roland Barthes} $\to$ \textbf{Bibliothèque nationale de France} $\to$ \textbf{Charles Baudelaire} $\to$ Georges Bataille \\
French & Pierre Klossowski $\to$ Jacques Lacan & 5 / 35 & Pierre Klossowski $\to$ \textbf{\textit{Esprit} (magazine)} $\to$ Jacques Lacan \\
French & Pierre Klossowski $\to$ Michel Foucault & 2 / 35 & Pierre Klossowski $\to$ \textbf{Guillaume Apollinaire} $\to$ \textbf{\textit{Éditions Gallimard}} $\to$ Michel Foucault \\
French & Pierre Klossowski $\to$ Friedrich Nietzsche & 7 / 35 & Pierre Klossowski $\to$ \textbf{Guy Debord} $\to$ \textbf{Jacques Ellul} $\to$ Friedrich Nietzsche \\
French & Pierre Klossowski $\to$ Friedrich Nietzsche & 15 / 35 & Pierre Klossowski $\to$ \textbf{\textit{Esprit} (magazine)} $\to$ \textbf{Jacques Lacan} $\to$ Friedrich Nietzsche \\
French & Pierre Klossowski $\to$ André Gide & 4 / 35 & Pierre Klossowski $\to$ \textbf{Lev Shestov} $\to$ André Gide \\
French & Pierre Klossowski $\to$ Marquis de Sade & 5 / 31 & Pierre Klossowski $\to$ \textbf{\textit{Esprit} (magazine)} $\to$ \textbf{May 68} $\to$ Marquis de Sade \\ \hline
German & Pierre Klossowski $\to$ Georges Bataille & 6 / 6 & Pierre Klossowski $\to$ Cologne $\to$ \textbf{Bildung} $\to$ Truth $\to$ \textbf{Karl Rahner} $\to$ Maurice Blondel $\to$ Henri Lefebvre $\to$ Georges Bataille \\
German & Pierre Klossowski $\to$ Friedrich Nietzsche & 1 / 40 & Pierre Klossowski $\to$ \textbf{Georg Wilhelm Friedrich Hegel} $\to$ Idea $\to$ Friedrich Nietzsche \\
German & Pierre Klossowski $\to$ André Gide & 3 / 16 & Pierre Klossowski $\to$ Cologne $\to$ \textbf{Nazi seizure of power} $\to$ Austria $\to$ Surrealism $\to$ André Gide \\
German & Pierre Klossowski $\to$ Marquis de Sade & 5 / 9 & Pierre Klossowski $\to$ Cologne $\to$ \textbf{Heinrich Heine} $\to$ \textbf{Dreyfus affair} $\to$ Geneviève Halévy $\to$ Marquis de Sade \\ \hline
English & Pierre Klossowski $\to$ Georges Bataille & 4 / 9 & Pierre Klossowski $\to$ \textbf{Brian Evenson} $\to$ Doctor of Philosophy $\to$ Bachelor's degree $\to$ Social science $\to$ Literary criticism $\to$ \textbf{Paul de Man} $\to$ Georges Bataille \\
English & Pierre Klossowski $\to$ Jacques Lacan & 1 / 9 & Pierre Klossowski $\to$ \textbf{Pier Paolo Pasolini} $\to$ \textbf{Intellectual} $\to$ \textbf{Existentialism} $\to$ Jacques Lacan \\
English & Pierre Klossowski $\to$ Michel Foucault & 1 / 9 & Pierre Klossowski $\to$ Szlachta $\to$ \textbf{London} $\to$ \textbf{\textit{Routledge}} $\to$ Michel Foucault \\
English & Pierre Klossowski $\to$ Friedrich Nietzsche & 4 / 9 & Pierre Klossowski $\to$ \textbf{Musée National d'Art Moderne} $\to$ Abstract art $\to$ Theodor W. Adorno $\to$ Friedrich Nietzsche \\
English & Pierre Klossowski $\to$ André Gide & 1 / 9 & Pierre Klossowski $\to$ \textbf{\textit{Salò, or the 120 Days of Sodom}} $\to$ Friedrich Nietzsche $\to$ \textbf{Thomas Carlyle} $\to$ André Gide \\
English & Pierre Klossowski $\to$ Marquis de Sade & 3 / 8 & Pierre Klossowski $\to$ \textbf{Pier Paolo Pasolini} $\to$ \textbf{\textit{The Guardian}} $\to$ \textbf{\textit{The Economist}} $\to$ Epistolary novel $\to$ Marquis de Sade \\
\hline
Japanese & Pierre Klossowski $\to$ Georges Bataille & 3 / 14 & Pierre Klossowski $\to$ \textbf{Shusaku Endo} $\to$ \textbf{Sei Ito} $\to$ French literature $\to$ Georges Bataille \\
Japanese & Pierre Klossowski $\to$ Jacques Lacan & 5 / 18 & Pierre Klossowski $\to$ \textbf{Akira Asada} $\to$ Jacques Lacan \\
Japanese & Pierre Klossowski $\to$ Michel Foucault & 2 / 21 & Pierre Klossowski $\to$ \textbf{Akira Asada} $\to$ Michel Foucault \\
Japanese & Pierre Klossowski $\to$ Friedrich Nietzsche & 8 / 21 & Pierre Klossowski $\to$ \textbf{Shusaku Endo} $\to$ \textbf{Tatsuo Hori} $\to$ \textbf{Hideo Kobayashi} $\to$ Friedrich Nietzsche \\
Japanese & Pierre Klossowski $\to$ Friedrich Nietzsche & 9 / 21 & Pierre Klossowski $\to$ 1930 $\to$ Japan $\to$ \textbf{Kenzaburo Oe} $\to$ \textbf{Kojin Karatani} $\to$ Friedrich Nietzsche \\

Japanese & Pierre Klossowski $\to$ Marquis de Sade & 8 / 14 & Pierre Klossowski $\to$ \textbf{\textit{Heibonsha}} $\to$ \textbf{\textit{Iwanami Shoten}} $\to$ \textbf{\textit{Kodansha}} $\to$ \textbf{Yukio Mishima} $\to$ \textbf{Kazushige Abe} $\to$ Marquis de Sade \\
 \hline
\end{tabularx}
\end{table}

\clearpage % Table 4 と Table 5 の間に元々ある改ページコマンド

\begin{table}[!ht]
\centering
\setstretch{1.19}
\scriptsize
\caption{Comparison of characteristic paths for Group 2 (Spivak-centered pairs).}
\label{tab:multilingual_paths_full}
\begin{tabularx}{\textwidth}{lll>{\raggedright\arraybackslash}X}
\hline
\textbf{Language} & \textbf{Source $\to$ Target} & \textbf{Rank/Total} & \textbf{Extracted Path} \\ \hline
French & Gayatri C. Spivak $\to$ Gilles Deleuze & 1 / 14 & Gayatri C. Spivak $\to$ Gilles Deleuze \\
French & Edward Said $\to$ Gayatri C. Spivak & 2 / 4 & Edward Said $\to$ \textbf{\textit{Éditions du Seuil}} $\to$ 2006 $\to$ \textbf{Socialist Party (France)} $\to$ \textbf{National Rally} $\to$ \textbf{Jean-Luc Mélenchon} $\to$ Proletarian internationalism $\to$ Gayatri C. Spivak \\
French & Gayatri C. Spivak $\to$ Judith Butler & 1 / 12 & Gayatri C. Spivak $\to$ \textbf{Paris 8 University} $\to$ \textbf{\textit{Le Monde}} $\to$ \textbf{Pierre Bourdieu} $\to$ Judith Butler \\
French & Gayatri C. Spivak $\to$ Jacques Derrida & 1 / 14 & Gayatri C. Spivak $\to$ \textbf{Michel Foucault} $\to$ \textbf{\textit{Éditions du Seuil}} $\to$ Jacques Derrida \\
French & Gayatri C. Spivak $\to$ Jacques Derrida & 6 / 14 & Gayatri C. Spivak $\to$ \textbf{Paris 8 University} $\to$ Science $\to$ Rhetoric $\to$ Jacques Derrida \\
French & Gayatri C. Spivak $\to$ Jacques Derrida & 7 / 14 & Gayatri C. Spivak $\to$ \textbf{Étienne Balibar} $\to$ \textbf{\textit{Presses Universitaires de France}} $\to$ Philosopher $\to$ Jacques Derrida \\
French & Jacques Derrida $\to$ Gayatri C. Spivak & 2 / 4 & Jacques Derrida $\to$ Algeria $\to$ \textbf{François Mitterrand} $\to$ \textbf{Jean-Luc Mélenchon} $\to$ Proletarian internationalism $\to$ Gayatri C. Spivak \\
 \hline
German & Edward Said $\to$ Gayatri C. Spivak & 2 / 6 & Edward Said $\to$ \textbf{Hannah Arendt} $\to$ Feminism $\to$ Gayatri C. Spivak \\
German & Edward Said $\to$ Gayatri C. Spivak & 6 / 6 & Edward Said $\to$ Discourse analysis $\to$ Subject $\to$ \textbf{Jürgen Habermas} $\to$ Cornell University $\to$ Gayatri C. Spivak \\
German & Gayatri C. Spivak $\to$ Jacques Derrida & 2 / 19 & Gayatri C. Spivak $\to$ Human rights $\to$ \textbf{Hannah Arendt} $\to$ Justice $\to$ Jacques Derrida \\
German & Gayatri C. Spivak $\to$ Jacques Derrida & 3 / 19 & Gayatri C. Spivak $\to$ Identity $\to$ \textbf{\textit{Suhrkamp Verlag}} $\to$ Jacques Derrida \\
German & Gayatri C. Spivak $\to$ Jacques Derrida & 5 / 19 & Gayatri C. Spivak $\to$ Reality $\to$ Subject $\to$ \textbf{Karl-Otto Apel} $\to$ Jacques Derrida \\
German & Gayatri C. Spivak $\to$ Jacques Derrida & 9 / 19 & Gayatri C. Spivak $\to$ British Academy $\to$ \textbf{Geisteswissenschaft} $\to$ Culture $\to$ Jacques Derrida \\
German & Gayatri C. Spivak $\to$ Jacques Derrida & 12 / 19 & Gayatri C. Spivak $\to$ Critical theory $\to$ Hannah Arendt $\to$ \textbf{Hanns Zischler} $\to$ Jacques Derrida \\
German & Gayatri C. Spivak $\to$ Jacques Derrida & 15 / 19 & Gayatri C. Spivak $\to$ Aphorism $\to$ \textbf{J.W. von Goethe} $\to$ Aristotle $\to$ God $\to$ Socrates $\to$ Jacques Derrida \\
German & Gayatri C. Spivak $\to$ Jacques Derrida & 19 / 19 & Gayatri C. Spivak $\to$ Columbia UP $\to$ Woodrow Wilson $\to$ \textbf{Otto von Bismarck} $\to$ \textbf{Ferdinand Lassalle} $\to$ Intellectual $\to$ Jacques Derrida \\
 \hline
English & Gayatri C. Spivak $\to$ Gilles Deleuze & 1 / 16 & Gayatri C. Spivak $\to$ \textbf{Continental philosophy} $\to$ Gilles Deleuze \\
English & Gayatri C. Spivak $\to$ Edward Said & 5 / 13 & Gayatri C. Spivak $\to$ \textbf{New York University} $\to$ Sociology $\to$ Literary criticism $\to$ Edward Said \\
English & Gayatri C. Spivak $\to$ Edward Said & 8 / 13 & Gayatri C. Spivak $\to$ \textbf{Terry Eagleton} $\to$ \textbf{\textit{The Guardian}} $\to$ \textbf{University of Cambridge} $\to$ \textbf{Salman Rushdie} $\to$ Edward Said \\
English & Gayatri C. Spivak $\to$ Edward Said & 10 / 13 & Gayatri C. Spivak $\to$ Contemporary philosophy $\to$ Modern philosophy $\to$ Friedrich Nietzsche $\to$ \textbf{Thomas Carlyle} $\to$ \textbf{Rudyard Kipling} $\to$ Edward Said \\
English & Gayatri C. Spivak $\to$ Jacques Derrida & 1 / 16 & Gayatri C. Spivak $\to$ \textbf{Other (philosophy)} $\to$ Jacques Derrida \\
English & Michel Foucault $\to$ Gayatri C. Spivak & 4 / 7 & Michel Foucault $\to$ Age of Enlightenment $\to$ Marxism $\to$ \textbf{New Left} $\to$ Gayatri C. Spivak \\ \hline
Japanese & Gilles Deleuze $\to$ Gayatri C. Spivak & 1 / 2 & Gilles Deleuze $\to$ \textbf{Akira Asada} $\to$ \textbf{Kojin Karatani} $\to$ Gayatri Chakravorty Spivak \\
Japanese & Edward Said $\to$ Gayatri C. Spivak & 1 / 2 & Edward Said $\to$ \textbf{Kenzaburo Oe} $\to$ \textbf{Kojin Karatani} $\to$ Gayatri C. Spivak \\
Japanese & Edward Said $\to$ Gayatri C. Spivak & 2 / 2 & Edward Said $\to$ \textbf{\textit{Iwanami Shoten}} $\to$ \textbf{\textit{Kodansha}} $\to$ 1990s $\to$ Gayatri C. Spivak \\
Japanese & Gayatri C. Spivak $\to$ Homi K. Bhabha & 1 / 1 & Gayatri C. Spivak $\to$ \textbf{\textit{Iwanami Shoten}} $\to$ \textbf{Takaaki Yoshimoto} $\to$ Jacques Derrida $\to$ Homi K. Bhabha \\
Japanese & Homi K. Bhabha $\to$ Gayatri C. Spivak & 1 / 2 & Homi K. Bhabha $\to$ Professor $\to$ Japan $\to$ \textbf{Kenzaburo Oe} $\to$ \textbf{Kojin Karatani} $\to$ Gayatri C. Spivak \\
Japanese & Gayatri C. Spivak $\to$ Judith Butler & 1 / 4 & Gayatri C. Spivak $\to$ \textbf{\textit{Iwanami Shoten}} $\to$ \textbf{\textit{Kodansha}} $\to$ 1990s $\to$ Judith Butler \\
Japanese & Gayatri C. Spivak $\to$ Jacques Derrida & 1 / 4 & Gayatri C. Spivak $\to$ \textbf{\textit{Iwanami Shoten}} $\to$ \textbf{Takaaki Yoshimoto} $\to$ Jacques Derrida \\
Japanese & Jacques Derrida $\to$ Gayatri C. Spivak & 1 / 2 & Jacques Derrida $\to$ \textbf{Kojin Karatani} $\to$ Gayatri C. Spivak \\ \hline
\end{tabularx}
\end{table}

\clearpage

\section{Discussion}

% --- Group 1 ---
As detailed in Table~\ref{tab:group1_paths}, while the extracted paths for Group 1 (Klossowski-centered pairs) represent only a specific subset of the overall network, they effectively illustrate how culture-specific contexts and concepts can emerge within each linguistic sphere. In the French edition, paths are heavily mediated by iconic domestic publishing institutions, cultural repositories, and avant-garde movements. Connections frequently route through major publishing houses such as \textit{Éditions Gallimard} (as seen in the lineage to Foucault via Apollinaire) and prominent intellectual journals like  \textit{Esprit} (which serves as a direct conduit to Lacan) and \textit{Les Temps modernes}. Furthermore, foundational cultural spaces like the ``Bibliothèque nationale de France'' (BnF) and highly specific socio-political milestones—such as the structural link from \textit{Esprit} to the Marquis de Sade via the events of ``May 68''—appear as crucial conduits. The French network also contextualizes Klossowski through unique local artistic lineages, including Guy Debord (Situationism), Antonin Artaud and Roland Barthes.

In the German edition, we can observe paths that incorporate prominent German thinkers and theologians (e.g., ``Georg Wilhelm Friedrich Hegel'', ``Heinrich Heine'', and ``Karl Rahner''), specialized cultural concepts such as ``Bildung'', and macro-historical events closely linked to Franco-German relations and the history of European anti-Semitism, such as ``the Nazi seizure of power'' and ``the Dreyfus affair''.

The English edition, on the other hand, features paths mediated by its own distinct cultural landscape. Connections appear through artistic and cultural products like the ``Mus\'{e}e National d'Art Moderne'' and Pasolini's film \textit{Sal\`{o}, or the 120 Days of Sodom}, academic pathways and institutional infrastructures (such as the route to Foucault via London and the prominent Anglophone publisher \textit{Routledge}), specialized theoretical categorizations (``Existentialism'' leading to Lacan), and major Anglophone media outlets like \textit{The Guardian} and \textit{The Economist}.

Similarly, the Japanese edition reveals its own unique cultural localization. Some characteristic paths feature domestic literary and critical figures, such as the lineage from ``Shusaku Endo'' to ``Tatsuo Hori'' and ``Hideo Kobayashi'', or via ``Kenzaburo Oe'' and ``Kojin Karatani''. Most notably, the paths to both Lacan and Foucault are structurally pinpointed through the prominent Japanese critic ``Akira Asada'', capturing the unique domestic landscape of New Academism that catalyzes the reception of French post-structuralism in Japan during the late 20th century. Furthermore, the connection to the Marquis de Sade traces a sequence of major domestic publishers (\textit{Heibonsha} $\to$ \textit{Iwanami Shoten} $\to$ \textit{Kodansha}) before reaching Japanese novelists. Rather than making a definitive structural claim about the entire network, these examples highlight how the Japanese Wikipedia incorporates its specific domestic literary history and publishing culture as part of the context for understanding French philosophy.

% --- Group 2 ---
To further demonstrate how local cultural frameworks contextualize these global figures, we compared the extracted paths for Group 2 (centered on Gayatri C. Spivak and post-colonial/post-structuralist thinkers). As summarized in Table~\ref{tab:multilingual_paths_full}, the intermediary nodes significantly differ, revealing the distinct intellectual landscapes of each linguistic sphere.

In the French edition, the network constructs an organic, contiguous landscape where post-structuralist thought is deeply bound to domestic academic centers and political apparatuses. Most notably, the French structure preserves a direct, length-1 exclusive link connecting Spivak to Deleuze, reflecting an unmediated relational proximity unique to the French Wikipedia. Paths from Spivak to Derrida and Butler repeatedly utilize localized hubs such as ``Paris 8 University (Vincennes)'', \textit{Éditions du Seuil}, and \textit{Presses Universitaires de France (PUF)}, alongside prominent French sociologists and feminists like ``Pierre Bourdieu'' and ``Christine Delphy''. Intriguingly, the French edition also explicitly maps post-structuralist trajectories onto domestic party politics and state authority, generating unique chains that pass through French presidents and contemporary political figures (e.g., ``François Mitterrand'', ``Jean-Luc Mélenchon'', the ``Socialist Party'', and the ``National Rally'').

In the German edition, paths often invoke domestic political philosophers, historical figures, and distinctively German academic concepts. For instance, connections from Spivak or Said frequently pass through Frankfurt School thinkers and political theorists like ``Hannah Arendt'', ``Karl-Otto Apel'' and ``J\"urgen Habermas'', or incorporate the uniquely German academic concept of ``Geisteswissenschaft'' (human sciences). The paths also feature major domestic publishers like \textit{Suhrkamp Verlag} and historical figures such as ``Otto von Bismarck'' and ``J.W. von Goethe'', reflecting how these international discourses are mapped onto familiar domestic reference points.

The English edition exhibits paths deeply anchored in Anglo-American academic institutions, literature, and media. For instance, the connection from Spivak to Deleuze is routed through a broad umbrella term, ``Continental philosophy''. Connections to Said feature prominent academic and media nodes, forming chains such as ``Terry Eagleton'' $\to$ \textit{The Guardian} $\to$ ``University of Cambridge'' $\to$ ``Salman Rushdie'', or passing through ``New York University''. It also contextualizes these thinkers through British Victorian writers like ``Thomas Carlyle'' and ``Rudyard Kipling'', as well as Anglophone political movements like the ``New Left''.

Consistent with the findings in Group 1, the Japanese edition reveals a structural reliance on domestic intellectual figures and institutional infrastructures. Paths connecting post-colonial and post-structuralist concepts pass through the prominent novelist-critic duo ``Kenzaburo Oe'' and ``Kojin Karatani''. This mediation, much like the patterns observed in Group 1, reflects the specific mode of intellectual reception within the domestic context. Furthermore, institutional nodes appear within the paths; connections linking Spivak to Butler and Bhabha pass through the major academic publishers \textit{Iwanami Shoten} and \textit{Kodansha}, or the critic ``Takaaki Yoshimoto''.

% --- 総括 ---
Ultimately, a structural contrast emerges in how these language editions contextualize global philosophical figures through their respective intermediary nodes. While the German and English editions tend to include nodes related to local history, academia, and media, the Japanese paths are characterized by the presence of domestic novelists (e.g., Oe, Mishima), critics (e.g., ``Kojin Karatani'', ``Takaaki Yoshimoto''), and publishers (e.g., \textit{Iwanami Shoten}, \textit{Kodansha}). This configuration reflects a context where international figures are linked through domestic actors and institutions associated with their reception. Conversely, paths in the French edition consistently incorporate corporate publishers, national libraries, specific universities, and active political parties. This indicates that within the French Wikipedia, philosophical connections are contextualized through various domestic institutional and political entities rather than being mediated primarily by individual critics or specific publishing histories.

\section{Limitations and future work}

While the proposed method effectively visualizes cultural divergences in knowledge structures, several limitations remain to be addressed in future research.

The networks analyzed in this study are derived from Wikipedia, a platform governed by collaborative editing. Consequently, the results may be influenced by the specific activities of editing communities or individual biases, rather than solely reflecting academic or professional consensus. In particular, the observed cross-lingual divergences might partly stem from platform-specific technical artifacts or community-specific behavioral variations rather than purely intellectual localization. These confounding factors include disparities in the size and activity patterns of editing communities, or differing local conventions regarding wikilink insertion—such as the customary high density of links to publishing houses in certain language editions. At the current stage, the proposed method treats the resulting graph topology as a given structural state and cannot strictly disentangle these platform-inherent variables from genuine cultural reception. Future studies should incorporate more formalized data sources, such as specialized encyclopedias or peer-reviewed bibliographic databases, to further validate the cross-domain applicability of this method.

The current definition of the ``Edge-Difference Graph'' employs a hard subtraction approach, where common edges are entirely removed. This can result in fragmented graphs or a significant loss of structural connectivity in regions where knowledge is highly shared across cultures. To overcome this, future iterations could implement soft edge reduction techniques, such as attenuating the weights of common edges rather than eliminating them, thereby preserving global context while still highlighting local uniqueness.

At present, the evaluation of the extracted paths relies on heuristic interpretations by researchers. To establish a more rigorous evaluative framework, it is necessary to compare the results with ground truth data, such as reference relations in professional dictionaries or expert-curated taxonomies. Furthermore, statistical validation through comparison with random graph models (e.g., null models) would help determine the significance of the discovered paths beyond mere coincidence.

Furthermore, the current framework uses a minimum-cost maximum-flow formulation with uniform edge capacities to guarantee that the extracted paths are edge-disjoint. However, this approach does not prevent node overlap. Consequently, influential intermediary concepts or actors, such as prominent critics or publishers, may still appear across multiple paths. While the recurrence of these central nodes reflects their historical importance within a specific cultural sphere, it can limit the thematic diversity of the paths. Future work should consider node-splitting techniques to find strictly node-disjoint paths, ensuring greater structural independence among the extracted trajectories.

Using Wikipedia data enables integration with Wikidata's structured metadata. Future research could combine node attributes such as entity type (e.g., person, organization, or location) with edge-difference analysis. This would provide a more quantitative and multi-faceted understanding of how specific categories or demographics influence the observed knowledge localization.

\section{Conclusion}

This study has demonstrated that combining Edge-Difference Graphs with cost-penalized network flow analysis is an effective approach for revealing localized cultural contexts within large-scale knowledge structures. Instead of relying on whole-graph statistics, our framework successfully extracted distinctive relational trajectories by bypassing generic hub nodes and highlighting culture-specific conceptual paths.

The empirical application to Wikipedia's networks provided concrete evidence of how global knowledge is uniquely structured across different linguistic spheres. Compared to the other editions, the French network shows a structure where philosophy is more closely connected to domestic publishing institutions, such as \textit{Éditions Gallimard} and the journal \textit{Esprit}, as well as academic and political organizations. In contrast, the German and English editions place these global figures within their own historical and academic contexts. The German network links them heavily to domestic intellectual traditions, such as the Frankfurt School and Hannah Arendt, and major historical events like the Nazi seizure of power. The English network structures concepts through prominent media outlets like \textit{The Guardian} and broad categories like Continental philosophy. Finally, the Japanese network reveals a unique reception history. Unlike the other editions, it localizes global ideas through specific domestic critics and writers, such as Akira Asada and Kojin Karatani, and major publishing houses like \textit{Iwanami Shoten} and \textit{Kodansha}.

The implications of this framework extend to fields such as digital humanities, cross-cultural information science, and knowledge graph engineering. Future research can enable more granular comparative analyses by incorporating node-disjoint path extraction and integrating metadata from Wikidata. Ultimately, this approach helps make cross-cultural knowledge localization systematically visible and interpretable.

\section*{Acknowledgement}
This work was supported by JSPS KAKENHI Grant Numbers 24K17472 and The Nippon Foundation HUMAI Research Project.

% --- ここから文献表（References） ---

\end{document}